\documentclass[aps,twocolumn,superscriptaddress,showpacs,letter,nofootinbib]{revtex4-1} 
\usepackage[utf8]{inputenc}
\usepackage{graphicx}% Include figure files
\usepackage{dcolumn}% Align table columns on decimal point
\usepackage{bm}% bold math
\usepackage{cancel}
\usepackage{color}             
\usepackage{epsfig}
\usepackage{amssymb}
\usepackage{amsmath}
\usepackage{units}
\usepackage{multirow}
\usepackage{hyperref}
\usepackage{cleveref}
\usepackage{lineno,color}
\usepackage{epstopdf}

\begin{document}
%\linenumbers

\newcommand{\ba}{\begin{array}}
\newcommand{\ea}{\end{array}}
\newcommand{\Pom}{I$\!\!$P}
\def\br{\begin{eqnarray}}
\def\er{\end{eqnarray}}
\def\be{\begin{equation}}
\def\ee{\end{equation}}
\def\a{\alpha}
\def\D{\Delta}
\def\g{\gamma}
\def\G{\Gamma}
\def\l{\lambda}
\def\L{\Lambda}
\def\n{\nu}
\def\({\left(}
\def\){\right)}
\def\s{\sigma}
\def\S{\Sigma}
\def\be{\begin{equation}}                                                    
\def\ee{\end{equation}}                                                    
\def\bea{\begin{eqnarray}}                                                    
\def\eea{\end{eqnarray}}
\def\b{\vec{b}}
\def\bb{\vec{b}'}     
\def\q{\vec{q}} 
\def\p{I\!\! P}
\def\pp{\vec{p}'_t}                                                    
\def\k{\vec{k}_t} 
\def\kk{\vec{k}'_t} 
\def\GeV{\rm GeV}
\def\slash#1{#1 \hskip-0.55em /}
\def\Pslash#1{#1 \hskip-0.6em /}
\def\sh{\hat s}
\def\sh2{{\hat s}^2}
\def\sd{$\sigma^{\rm D}_{{\rm low}M}~$}
\def\se{$\sigma_{\rm el}~$}

\renewcommand{\vec}[1]{\boldsymbol{#1}}
\newcommand{\dif}{\mathrm{d}}
\newcommand{\diff}[1]{\frac{\mathrm{d}#1}{#1}}
\newcommand{\xB}{x_{\scriptscriptstyle{B}}}
\newcommand{\slashed}[1]{\makebox[0pt][l]{/}#1}

%\twocolumn[\hsize\textwidth\columnwidth\hsize\csname %%% TWO COLUMN
%@twocolumnfalse\endcsname                            %%% TWO COLUMN

\title{Central multiplicity distributions in the multi-channel eikonal model.}

\author{E.~G.~S.~Luna}
\email{luna@if.ufrgs.br}
\affiliation{Instituto de F\'isica, Universidade Federal do Rio Grande do Sul, Caixa Postal 15051, 91501-970, Porto Alegre, RS, Brazil}
\author{M.~G.~Ryskin}
\email{ryskin@thd.pnpi.spb.ru}
\affiliation{NRC ``Kurchatov Institute'' - PNPI, \\Gatchina, 188300, Russia}

%\date{\today}

\begin{abstract}

The multiplicity distribution of charged particles produced in the central rapidity region ($|\eta|<2.5$) is calculated within the multi-channel eikonal model using the AGK cutting rules and compared with ATLAS data at $\sqrt{s}=7$ and 13 TeV. The effects of color reconnection and/or string percolation are discussed.

\end{abstract}

%\pacs{12.38.Lg, 13.85.Dz, 13.85.Lg}

\maketitle

\section{Introduction}

The study of the high-energy behavior of the $pp$ scattering amplitude is one of the goals of the LHC. The observed growth of the total cross section with energy is usually described in terms of Pomeron exchange. One may take either the QCD Pomeron or the soft Pomeron with the intercept $\alpha_{\Bbb P}(0)>1$ as the seed (Born) amplitude. 
However, at asymptotically high energies, $\sqrt s\to\infty$, the power-like growth of this (one Pomeron exchange) Born amplitude violates the Froissart limit $\sigma<C\cdot\ln^2s$~\cite{Fr}. To satisfy unitarity, the amplitude ${\cal A}(s,t)$ must therefore be unitarized. 
 
There are two main possibilities for implementing unitarization: the eikonal scheme and the $U$-matrix approach. It is not easy to determine which of these schemes is actually realized in Nature.

In Refs.~\cite{Cud,LMP,LMP2}, collider data ($\sqrt{s}\geq 200$ GeV) on the total cross section, $\sigma_{tot}$, the differential elastic cross section $d\sigma_{el}/dt$ at low momentum transfer ($|t|<0.1$ GeV$^2$), and the ratio $\rho=\mathrm{Re}/\mathrm{Im}$ at $t=0$ were analyzed in terms of both the eikonal and the $U$-matrix unitarization schemes. It was found that the available $\sigma_{tot}$, $d\sigma_{el}/dt$, and $\rho$ data are not sufficient to discriminate between the $U$-matrix and the eikonal approaches.

To address this issue, it was proposed in Ref.~\cite{agk1} to study the multiplicity distribution of secondaries produced in the central rapidity region using the AGK Reggeon cutting rules~\cite{AGK}.

It was shown that the $U$-matrix approach can be self-consistent (i.e., compatible with unitarity) in the case where only {\em one} Pomeron is allowed to be cut (that is, to produce secondaries). However, in this scenario, the probability of events with large multiplicity becomes too small. Thus, the $U$-matrix prediction disagrees with experiment, whereas a simple one-channel eikonal model describes the LHC data~\cite{atl-epg} quite well.

A limitation of the one-channel eikonal model is that it neglects the possibility of diffractive dissociation, whose contribution is not negligible.

In the present paper, we include diffractive dissociation using the Good–Walker formalism~\cite{GW}. Based on the AGK Reggeon cutting rules~\cite{AGK}, we analyze the multiplicity distributions generated by the {\em multi}-channel eikonal.

It was emphasized in \cite{FN} that, at high energies, the multiplicity distribution exhibits a shoulder structure in the region of large multiplicities. In \cite{FN} this shoulder was described using the two component (pure soft and semihard) model. Here we show that a similar shoulder is generated by the multi-channel eikonal due to the presence of the Good-Walker components with a large and a small cross sections (the component with a large cross section creates a larger number of cut Pomerons and therefore a larger multiplicity).
Moreover, in our case, the shoulder becomes even too pronounced, and it becomes necessary to reduce the number of secondaries produced when many Pomerons are present. Such a reduction may be due to color reconnection \cite{CR1,CR2,CR3} or string percolation effects \cite{BP,BP2,BP3}.

It is useful to clarify the relation between the present analytic treatment and the more detailed descriptions implemented in Monte Carlo event generators. In general-purpose generators, the effects associated with multiple elementary interactions are usually described in terms of multiple parton interactions, supplemented by parton showers, energy-momentum conservation, resonance decays, hadronization, and phenomenological mechanisms such as color reconnection. In generators based more directly on the Gribov-Regge framework, such as QGSJET and EPOS, the relation between multiple elementary interactions, multi-Pomeron exchange, and the corresponding cut diagrams is more explicit.

The purpose of the present approach is complementary to that of a full event generator. We do not aim to provide an exclusive event-by-event description of the final state. Instead, the AGK rules allow the contributions associated with different numbers of cut Pomerons to remain explicit, while the Good-Walker decomposition relates fluctuations in the interaction strength to different diffractive components of the proton wave function. This makes it possible to identify directly how these fluctuations generate the shoulder of the multiplicity distribution and why the assumption of independent particle production by each cut Pomeron becomes inadequate in configurations with a large Pomeron density.

To place the results in a broader phenomenological context, we compare our predictions at $\sqrt{s}=7$ and 13 TeV with two complementary descriptions. The first is a weighted superposition of three negative binomial distributions, fitted directly to the measured multiplicity spectra. The second is the QGSJET-II-04 Monte Carlo result, which includes a more complete treatment of multiple scattering, diffraction, energy-momentum conservation, and hadronization. These comparisons help to distinguish the analytic transparency of the present framework from the greater event-level completeness of a Monte Carlo simulation.

The outline of this paper is as follows. In Sec.~II we briefly review the Good–Walker formalism. In Sec.~III we remind the AGK rules and present the corresponding formulae for the multi-channel eikonal case. In Sec.~IV we calculate the multiplicity distributions of secondaries produced in the central rapidity region, assuming that the particle multiplicity produced by a single cut Pomeron follows a Poisson distribution.
The results are compared with LHC data at $\sqrt{s}=7$ TeV and at the highest currently available energy, $\sqrt{s}=13$ TeV. In Sec.~V we discuss the effects that may reduce the number of secondaries in events with a very large Pomeron density. Our conclusions are presented in Sec.~VI.

\section{Good-Walker formalism}

Diffractive dissociation is a consequence of the {\em internal structure} of hadrons. At high energies, it is most conveniently described in the impact-parameter representation, $b$. In this regime, the value of $b$ does not change during the interaction, and the lifetime of each particular Fock component of the incoming hadron wave function, i.e. the hadronic fluctuations, is large, $\tau \sim E/m^2$. Therefore, during this time interval, the corresponding Fock state may be considered as effectively ``frozen''.

Each Fock component can undergo scattering independently, thereby destroying the coherence of the fluctuations. As a result, the outgoing superposition of states differs from the incident particle and will most likely contain multiparticle states. Thus, in addition to elastic scattering, {\em inelastic} processes may also occur.

To calculate diffractive dissociation, one may enlarge the set of intermediate states $(p,N^{*}_{a})$ beyond the single elastic channel and introduce a multichannel eikonal. However, it is more convenient to follow Good and Walker \cite{GW} and introduce states $\phi_{k}$ that diagonalize the $T$-matrix (which, for example, in the proton case, describes transitions such as $p\to N^{*}$ and $N^{*}_a\to N^{*}_b$). These eigenstates undergo only elastic scattering, and there are no off-diagonal transitions,
\begin{eqnarray}
\langle \phi_i|T|\phi_k\rangle = 0\qquad{\rm for}\ i\neq k.
\end{eqnarray}
Thus, a state $k$ cannot diffractively dissociate into a state $i$. In terms of the Good–Walker eigenstates $\phi_{i}$, for each state we have a simple one-channel eikonal.
We denote by $a$ the orthogonal matrix that diagonalizes ${\rm Im}\,T$, such that
\begin{eqnarray}
{\rm Im}\,T \; = \; aFa^T \quad\quad {\rm with}
\quad\quad \langle \phi_i |F| \phi_k \rangle \; = \; F_k \:
\delta_{ik},
\label{eq:a3}
\end{eqnarray}
where $F_k$ is the probability amplitude for the hadronic process to proceed via the diffractive eigenstate $\phi_k$.
Now consider the diffractive dissociation of an incoming state $|i\rangle$. It can be written as a superposition of the eigenstates $\phi_{k}$,
\begin{eqnarray}
| i \rangle \; = \; \sum_k \: a_{ik} \: | \phi_{k}
\rangle.
\label{eq:a4}
\end{eqnarray}

The elastic scattering amplitude then satisfies
\begin{eqnarray}
\langle i |{\rm Im}~T| i \rangle \; = \; \sum_k
\: |a_{ik}|^2 \: F_k \; = \; \langle F \rangle,
\label{eq:a5}
\end{eqnarray}
where $F_{k} \equiv \langle \phi_k |F| \phi_k \rangle$, and where the brackets $\langle F \rangle$ denote the average of $F$ over the initial probability distribution of diffractive eigenstates.
After the diffractive scattering described by $T_{fi}$, the final state $| f \rangle$ will, in general, be a different superposition of eigenstates from that of the initial state $| i \rangle$ given in (\ref{eq:a4}). Neglecting the real parts of the diffractive amplitudes, the cross sections at a given impact parameter $b$ read
\begin{eqnarray}
\frac{d \sigma_{\rm tot}}{d^2 b} & = & 2 \:
{\rm Im} \langle i |T| i \rangle  =  2 \: \sum_k
\: |a_{ik}|^2 \: F_k  =  2\, \langle F \rangle , \label{eq:b1} \\
& & \nonumber\\
\frac{d \sigma_{\rm el}}{d^2 b} & = & \left | \langle i |T| i
\rangle \right |^2  =  \left |
\sum_k \: |a_{ik}|^2 \: F_k \right |^2  =  \langle F \rangle^2 , \label{eq:b2}\\
& & \nonumber \\
\frac{d \sigma_{\rm el  + SD}}{d^2 b} & = & \sum_k \: \left |
\langle \phi_k |T| i \rangle \right |^2  =  \sum_k \:
|a_{ik}|^2 \: F_k^2  =  \langle F^2 \rangle .
\label{eq:b3}
\end{eqnarray}
It follows that the cross section for single diffractive dissociation of a proton is
\begin{eqnarray}
\frac{d \sigma_{\rm SD}}{d^2 b} \; = \; \langle
F^2 \rangle \: - \: \langle F \rangle^2,
\label{eq:a7}
\end{eqnarray}
which is given by the statistical dispersion of the absorption probabilities of the diffractive eigenstates. Here, the average is taken over the components $k$ of the incoming proton that undergoes dissociation. If the averages are taken over the components of both incoming particles, then (\ref{eq:a7}) corresponds to the sum of the single and double diffractive dissociation cross sections.

The Good-Walker eigenstates should be understood as effective representations of long-lived configurations of the proton wave function, with different absorption probabilities. In a microscopic QCD description, such fluctuations may differ in their partonic content, transverse size, and gluon density. However, the Good-Walker formalism itself does not establish a unique correspondence between an individual diffractive eigenstate and a particular quark-gluon Fock component. The eigenstates introduced here, therefore, provide a phenomenological basis in which fluctuations of the internal proton structure are characterized by their different interaction strengths.

In principle, the proton wave function may contain a continuous distribution, or a very large set, of configurations with different absorption probabilities. A model with a finite number of Good-Walker channels should consequently be regarded as a discrete approximation to this underlying distribution. The two-channel model is the minimal nontrivial realization: unlike a single-channel description, it allows a nonzero dispersion of interaction strengths and hence low-mass diffractive dissociation. The five-channel model provides a finer discretization of the same distribution and is used to test the stability of the results with respect to the number of effective configurations retained. Thus, the two- and five-channel cases do not represent different dynamical mechanisms or parametrizations with two and five free parameters, respectively, but rather two approximations of different resolutions to the same Good-Walker picture.

As shown by Eq. (\ref{eq:a7}), the low-mass diffractive dissociation cross section is controlled by the dispersion of the absorption amplitudes over the Good--Walker components. In the explicit implementation introduced in Sec .~IV, this dispersion is parametrized by the relative Pomeron couplings $\gamma_i$ of the diffractive eigenstates. Their weighted average fixes the mean interaction strength, whereas their spread determines the magnitude of diffractive dissociation.

\section{AGK rules}

According to the AGK rules \cite{AGK}, the contribution of a diagram with $n$ Pomerons to the multiparticle production cross section is obtained by cutting one or several ($k$) Pomerons. Each cut Pomeron produces a number of secondaries that are uniformly distributed in rapidity space. In this case the forward amplitude reads
\begin{eqnarray}
\textnormal{Im}{\cal A}(s,t=0)=s\sum_n C_n(-1)^{n-1} {\Bbb P}^n .
\end{eqnarray}  

In the impact-parameter representation the function ${\Bbb P}(s,b)$ represents the opacity ${\rm Re}\,\Omega = {\Bbb P}$, given by 
\begin{eqnarray}
\Omega(s,b)=-\frac{2i}s\int_0^{\infty} \!\! q\, dq\, J_0(qb)\, {\cal A}^{one\, \Bbb P}(s,t=-q^2),
\end{eqnarray}
where $J_0$ is the Bessel function.
By cutting $k$ Pomerons from the term $C_n(-1)^{n-1}{\Bbb P}^n$, the resulting cross section is 
$\sigma_n^k=2C_nc^k_n{\Bbb P}^n$,
where
\begin{eqnarray}
\label{agk}
c^{k\neq 0}_n = (-1)^{n-k}2^{n-1}\frac{n!}{k!(n-k)!},
\end{eqnarray}
\begin{eqnarray}
c^{k=0}_n=(-1)^n(2^{n-1}-1).
\end{eqnarray}

The AGK factor $c^{k=0}_{n}$ corresponds to the cut between the Pomerons when no secondaries particles are produced in the central region.\footnote{For a cut Pomeron, we consider the discontinuity ($disc=2\mbox{Im}{\cal A}$), while an uncut Pomeron may lie either to the left or to the right of the cut (which gives rise to the factor 2 in (\ref{agk})). In this way, the real part cancels.}$^,$
\footnote{Using the identity $\sum_{k=0}^n(-1)^kn!/(k! (n-k)!)=0$, it is easy to verify that the sum over all possible cuts, $\frac{1}{2}\sum_k\sigma^{k}_{n}=C_n(-1)^{n-1}{\Bbb P}^n$, is equal to the $n$-Pomeron exchange contribution to the total cross section. Strictly speaking, the first term, $2^{n-1}$ in $c^{k=0}_{n}$, should be multiplied by ${\Bbb P}$ (i.e., by the imaginary part of the one-Pomeron exchange amplitude). In contrast, in the last term, $1$, corresponding to the cut lying to the left (or to the right) of all the Pomerons, the full complex amplitude must be retained.}
Thus, for a process with multiplicity $k$ times larger than that produced by a single Pomeron, the cross section reads
\begin{eqnarray}
\label{k-cut}
\sigma^k (s,b) = 2\sum_nC_n\cdot (-1)^{n-k}2^{n-1}\frac{n![{\Bbb P}(s,b)]^n}{k!(n-k)!} .
\end{eqnarray}

Note that, since the expansion in powers of ${\Bbb P}$ has alternating signs, the  contribution of $k$ cut Pomerons coming from the exchange of $n$ Pomerons may be negative (see the factor $(-1)^n$ in (\ref{agk})). In such a case, it describes a screening correction to the process, with $k$ cut Pomerons generated by terms with larger $n$.

The factor $n!{\Bbb P}^n/(n-k)!$ in (\ref{k-cut}) can be written as the $k$-th derivative with respect to ${\Bbb P}$,
\begin{eqnarray}
\label{der}
\frac{n!{\Bbb P}^n}{(n-k)!} = {\Bbb P}^k\left(\frac{d}{d{\Bbb P}}\right)^k {\Bbb P}^n .
\end{eqnarray}

Thus, substituting $C_{n}=C_{n}^{eik}=2^{-n}/n!$ into (\ref{k-cut}), the cross sections $\sigma^k(s)=\int d^{2}b\, \sigma^k(s,b)$ for $k\geq 1$ take the form
\begin{eqnarray}
\label{eik-s}
\sigma^k_{eik}(s) = \int d^2b\, \frac{[\mbox{Re}\,\Omega(s,b)]^k}{k!} 
\exp(-\mbox{Re}\,\Omega(s,b)) . 
\end{eqnarray}
In the multichannel case, one must consider the collision between each pair of components ($i$, $j$) separately. Accordingly, the expression (\ref{eik-s}) should be replaced by
\begin{eqnarray}
\label{eik-sn}
\sigma^k_{eik}(s) & = & \sum_{i,j=1}^{n_{channel}}\int d^2b|a_ia_j|^2 \nonumber \\
& & \times \left( \frac{[\mbox{Re}\,\Omega_{ij}(s,b)]^k}{k!} 
\exp(-\mbox{Re}\,\Omega_{ij}(s,b))\right)\!,
\end{eqnarray}
where $n_{channel}$ denotes the number of Good-Walker components and $\Omega_{ij}$ is the opacity corresponding to the collision of the eigenstates $i$ and $j$. In terms of these eigenstates, the proton wave  function reads $|p\rangle = \sum_i a_i|\phi_i\rangle $, and the $pp$ cross sections are:
\begin{eqnarray}
\sigma_{tot}=2\sum_{i,j}\int d^2b|a_ia_j|^2\mbox{Im}\left(1-e^{-\Omega_{ij}(s,b)/2}\right)  ,
\end{eqnarray}

\begin{eqnarray}
\sigma_{el}=\int d^2b\left|\sum_{i,j}|a_ia_j|^2\left(1-e^{-\Omega_{ij}(s,b)/2}\right)\right|^2  ,
\end{eqnarray}
and
\begin{eqnarray}
\sigma_{el}+\sigma^D=\sum_{i,j}\int d^2b|a_ia_j|^2\left|1-e^{-\Omega_{ij}(s,b)/2}\right|^2  ,
\end{eqnarray}
where $\sigma^D$ includes the dissociation of both (the beam and the target) protons plus the dissociation of both incoming nucleons simultaneously.

\section{Numerical estimates}

In order to determine the Pomeron parameters, or equivalently the opacity function $\Omega(s,b)$ describing the interaction, we performed a global fit to the total cross section data for $pp$ and $\bar{p}p$ scattering above $\sqrt{s}= 10$ GeV \cite{pdg}, as well as to the $pp$ differential cross section at $\sqrt{s}=$ 7 \cite{atlas7Tev}, 8 \cite{atlas8Tev}, and 13 \cite{atlas13Tev} TeV in the range $|t|_{min} \leq |t|\leq 0.1$ GeV$^{2}$. Here, $|t|_{min} \sim 10 |t|_{int} = 0.71/\sigma_{tot}$, ensuring that the selected region is dominated by nuclear scattering \cite{luna004}. For energies above the ISR range ($\sqrt{s}\gtrsim 63$ GeV), we used $pp$ scattering data exclusively from the ATLAS Collaboration.

We emphasize that the elastic parametrization adopted here is restricted to the forward region, $|t|_{\min} \leq |t| \leq 0.1$ GeV$^{2}$, and should not be extrapolated quantitatively to the dip-bump region. In the present implementation, all Good-Walker eigenstates share the same transverse profile and differ only in their coupling strengths. Increasing the number of components from two to five offers a finer discretization of interaction strengths but does not add transverse scales, zeros, or relative phases in the elastic amplitude. A realistic description of the diffractive dip and subsequent bump would require a substantially more detailed elastic-scattering model with channel-dependent form factors, a flexible nonexponential $t$-dependence, careful treatment of the real part of the amplitude, and possibly an odd-signature contribution such as the Odderon. Since the elastic analysis performed here is intended solely to determine the forward opacity entering the multiplicity calculation, the study of the larger-$|t|$ region lies beyond the scope of the present work.

The single-Pomeron exchange contribution is given by
\begin{eqnarray}
{\cal A}_{\Bbb P}(s,t) =  \eta_{\Bbb P}(t)\, \beta_{\Bbb P}^{2}(t)\, \left( \frac{s}{s_{0}} \right)^{\alpha_{\Bbb P}(t)} ,
\end{eqnarray}
where $\eta_{\Bbb P}(t) = -e^{-i\frac{\pi}{2}\alpha_{\Bbb P}(t)}$ is the even signature factor. We assume an exponential form for the elastic proton-Pomeron vertex, $\beta_{\Bbb P}(t)=\beta_{\Bbb P}(0)\, e^{r_{\Bbb P}t/2}$, and adopt the simplest parametrization of the Pomeron trajectory, $\alpha_{\Bbb P}(t) = 1 + \epsilon + \alpha^{\prime}_{\Bbb P} t$. This leads to\,\footnote{The total opacity includes, in addition to the Pomeron contribution, secondary Reggeon terms, which account for the differences between the $pp$ and $\bar{p}p$ channels at low energies. However, at LHC energies, the Pomeron opacity dominates, and we therefore restrict our discussion to this contribution.}
\begin{eqnarray}
\Omega_{0}(s,b)=-i\eta_{\Bbb P}(0)\beta_{\Bbb P}^{2}(0)\left(\frac{s}{s_{0}}\right)^\epsilon\frac{1}{B}\exp\left(-\frac{b^{2}}{4B}\right) ,
\end{eqnarray}
where the slope parameter is given by
\begin{eqnarray}
B=r_{\Bbb P}+\alpha'_{\Bbb P}\left[ \ln \left( \frac{s}{s_0} \right) -i\frac \pi 2 \right] .
\end{eqnarray}

\begin{table*}
\centering
\caption{The values of the Pomeron parameters obtained from fits to elastic $pp$ and $\bar{p}p$ data \cite{pdg,atlas7Tev,atlas8Tev,atlas13Tev} and the corresponding values of the $\sigma^{D}$, $\sigma_{tot}$, and $\sigma_{el}$ at $\sqrt{s}=13$ TeV.}
\begin{ruledtabular}
\begin{tabular}{cccccc}
% \toprule
% \hline
  & \multicolumn{2}{c}{2 channels}  & \multicolumn{2}{c}{5 channels} \\
\cline{2-3} \cline{4-5}
 & $\gamma_{1}=1.545$, & $\gamma_{1}=1.786$, & $\gamma_{1}=1.73$, $\gamma_{2}=1.295$, $\gamma_{3}=1.0$, & $\gamma_{1}=1.93$, $\gamma_{2}=1.40$, $\gamma_{3}=1.0$,   \\
 & $\gamma_{2}=0.455$ & $\gamma_{2}=0.214$ & $\gamma_{4}=0.705$, $\gamma_{5}=0.27$ & $\gamma_{4}=0.60$, $\gamma_{5}=0.07$    \\
% & $\sigma^{D}=6.1$ mb & $\sigma^{D}=11.1$ & \hspace{2.5truecm} $\sigma^{D}=5.0$  \\
\hline
$\epsilon$ & 0.1220$\pm$0.0059 & 0.1542$\pm$0.0042 & 0.1208$\pm$0.0016 & 0.1588$\pm$0.0027  \\
$\alpha^{\prime}_{I\!\!P}$ (GeV$^{-2}$) & 0.185$\pm$0.018 & 0.108$\pm$0.012 & 0.1922$\pm$0.0011 & 0.088$\pm$0.027   \\
$\beta_{\Bbb P}(0)$ & 2.13$\pm$0.12 & 2.216$\pm$0.085 & 2.107$\pm$0.032 & 1.874$\pm$0.038  \\
$r_{\Bbb P}$ (GeV$^{-2}$) & 3.28$\pm$0.32 & 3.37$\pm$0.22 & 3.237$\pm$0.086 & 3.95$\pm$0.40   \\
\hline
$\sigma^{D}$ (mb) & 5.0 & 10.0 & 5.0 & 10.0 \\
$\sigma_{tot}$ (mb) & 103.5 & 103.3 & 103.4 & 102.7 \\
$\sigma_{el}$ (mb) & 26.8 & 26.6 & 26.8 & 26.6 \\
\hline
$\nu$ & 165 & 165 & 165 & 165 \\
\hline
$\chi^{2}/\nu$ & 0.77 & 0.85 & 0.77 & 0.88 \\
% \hline
\end{tabular}
\end{ruledtabular}
\label{tab001}
\end{table*}

\begin{figure*}[t]
\begin{center}
\vspace{-3.5cm}
\includegraphics[scale=0.6]{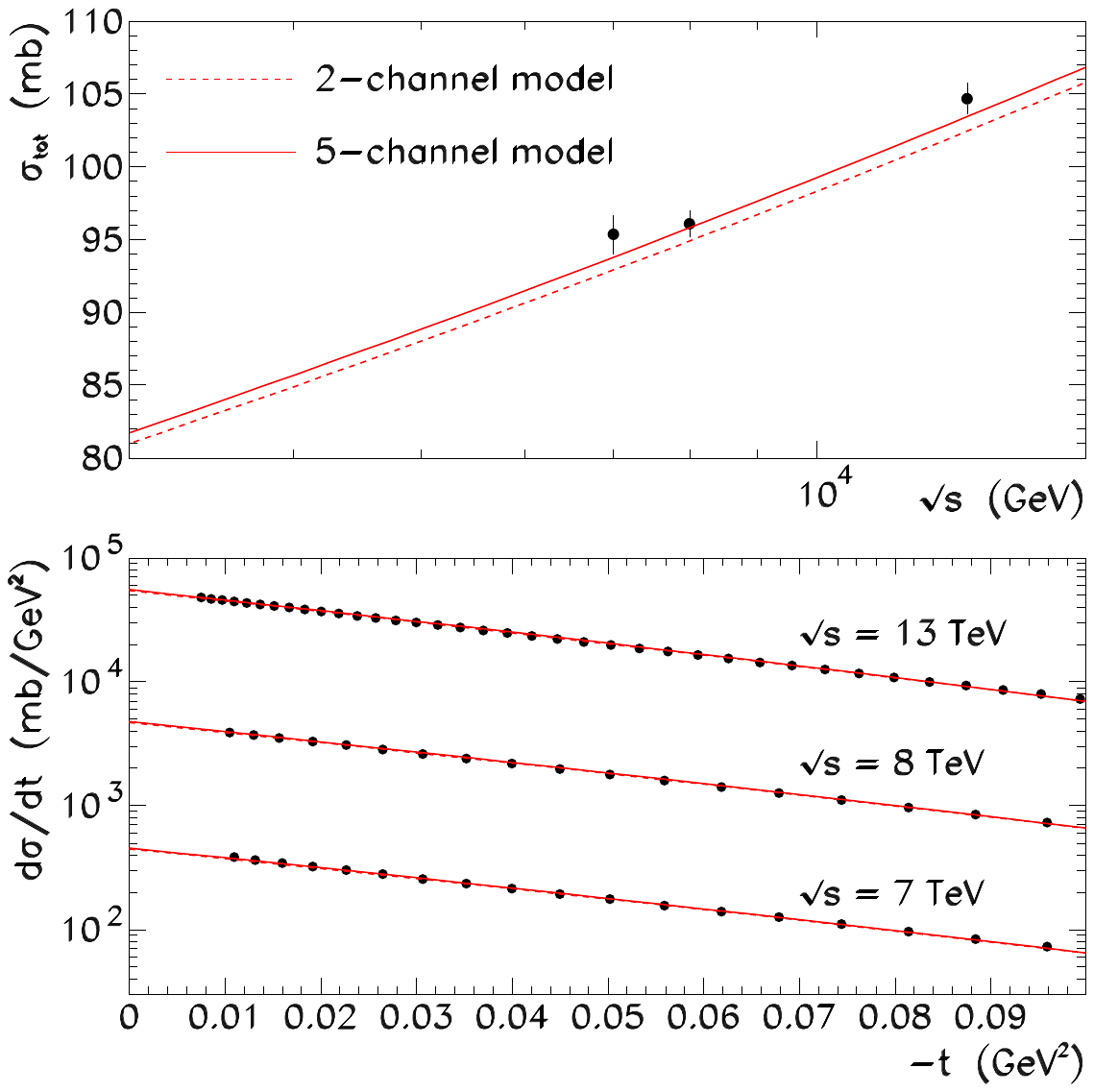}
%\hspace{-0.5cm}
%\includegraphics[scale=0.30]{u-3.pdf}
\vspace{-0.5cm}
\caption{\sf Total cross section (upper panel) \cite{pdg} and differential cross sections (lower panel) at $\sqrt{s}=$ 7, 8, and 13 TeV \cite{atlas7Tev,atlas8Tev,atlas13Tev}. The data in both panels are from the ATLAS Collaboration.
}
\label{f1}
\end{center}
\end{figure*}

To reduce the number of free parameters, we assume that all Good-Walker eigenstates have the same shape (all $a_{i}=1/\sqrt{n_{channel}}$), while allowing their couplings to differ, $\beta_i=\gamma_i\beta_{\Bbb P}(0)$. In this case, the opacity factorizes as $\Omega_{ij}(s,b)=\gamma_i\gamma_j\Omega_0(s,b)$. This parametrization provides a good description of the high-energy collider data on total and differential cross sections, as illustrated in the Fig.~\ref{f1} for the two- and five-channel eikonal models.
The Pomeron parameters, determined from a global fit to the data using a $\chi^{2}$ minimization procedure, where $\chi^{2}_{min}$ follows a $\chi^{2}$ distribution with $\nu$ degrees of freedom, are presented in Table~I, yielding $\chi^{2}/\nu = 0.77$ for both the two- and five-channel models. These values correspond to the fits in which the coefficients $\gamma_{i}$ are constrained such that the low-mass dissociation cross section satisfies $\sigma^{D} = 5$ mb.

By varying the parameters $\gamma_i$, one controls the dispersion of the Good–Walker eigenstate couplings and, consequently, the probability of diffractive dissociation.

To model particle production in the central region from a single Pomeron exchange, we start by assuming a Poisson distribution. This would be appropriate if particles were produced independently. However, in practice, short-range (in rapidity) correlations play an important role.
A first source of correlations arises from electric charge conservation, which implies that secondary particles are predominantly produced in oppositely charged pairs, such as $\pi^+$ and $\pi^-$. These pairs are typically separated by a relatively small rapidity interval. For this reason, it is more natural to consider a Poisson distribution in the number of pairs, $N^{{\Bbb P}}_{ch}/2$, where $N^{{\Bbb P}}_{ch}$ is the number of charged particles produced by a single Pomeron.
A second source of correlations is related to the production of intermediate objects such as resonances or mini-jets, which subsequently decay into several particles. If we denote by $C$ the average number of charged particles produced per such cluster, then the relevant Poisson variable becomes the number of clusters, $N^{{\Bbb P}}_{ch}/C$. Since each cluster must, on average, produce at least a pair of oppositely charged particles, one expects $C > 2$.

\begin{figure*}[t]
\begin{center}
\vspace{-3.5cm}
%\hspace{0.9cm}
\includegraphics[scale=0.6]{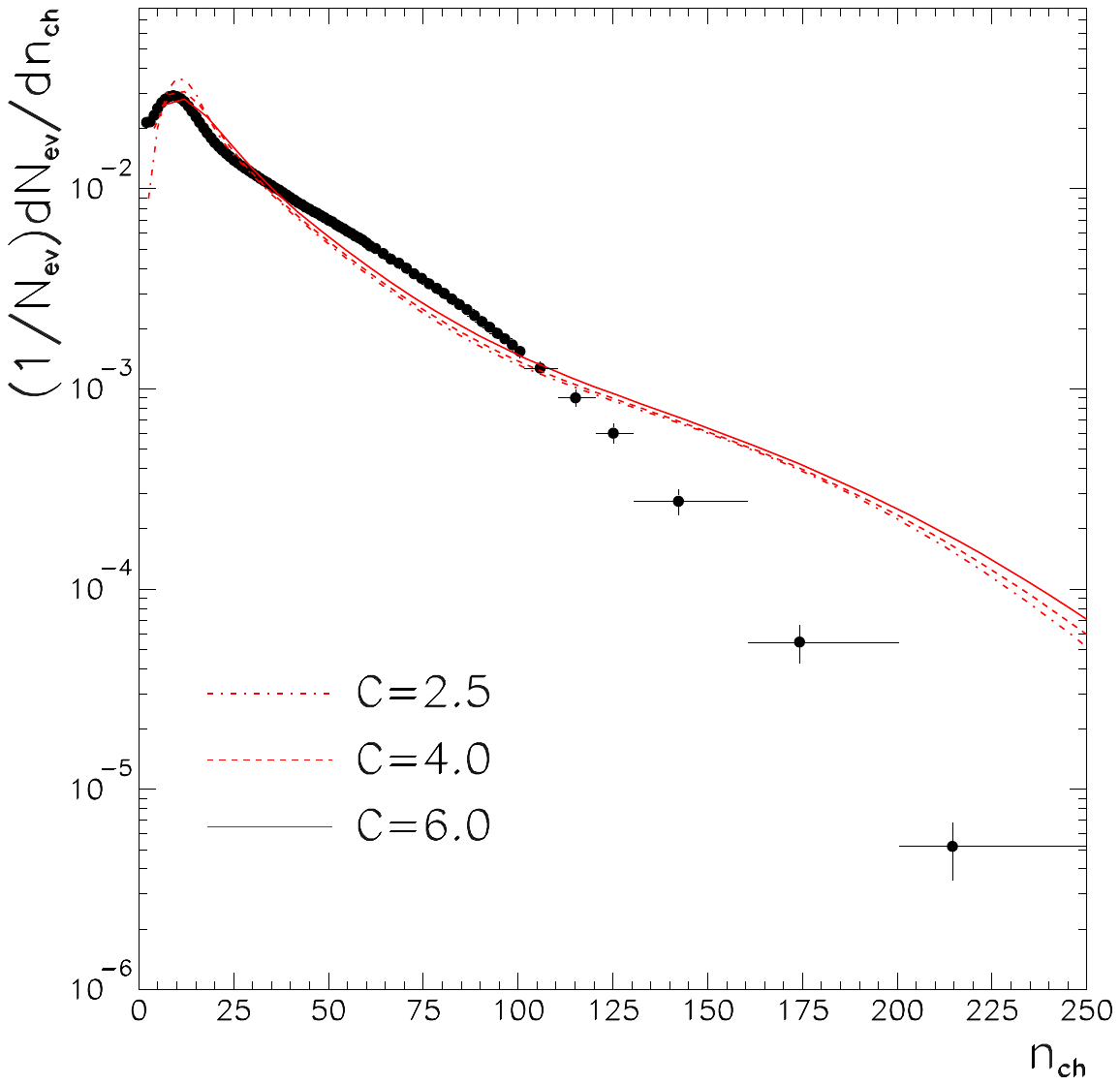}
\vspace{-0.5cm}
\caption{\sf Charged-particle multiplicity distribution in the central region ($-2.5<\eta<2.5$) for the two-channel eikonal case with different cluster multiplicities: $C$=2.5 (dash-dotted curve), $C=4$ (dashed curve) and $C=6$ (solid curve). Here, we take $\gamma_1=1.545$ and $\gamma_2=0.455$, leading to a dissociation cross section $\sigma^D=5.0$ mb. The data are from~\cite{atl-epg}.
}
\label{f2}
\end{center}
\end{figure*}

The results obtained within the two-channel eikonal unitarization are shown in Fig.~\ref{f2} for the values $C = 2.5$, $4$, and $6$. The parameter $N^{{\Bbb P}}_{ch}$ is fixed to reproduce the measured charged-particle density $dN_{ch}/d\eta$ reported in Ref.~\cite{atl-epg}. We observe that, except in the regions of very low or very high multiplicity, the dependence on the parameter $C$ is rather weak. This indicates that the overall shape of the multiplicity distribution is not primarily determined by the details of cluster formation. Instead, it is largely driven by fluctuations in the number of cut Pomerons, $k$.

\begin{figure*}[t]
\begin{center}
\vspace{-3.5cm}
\includegraphics[scale=0.6]{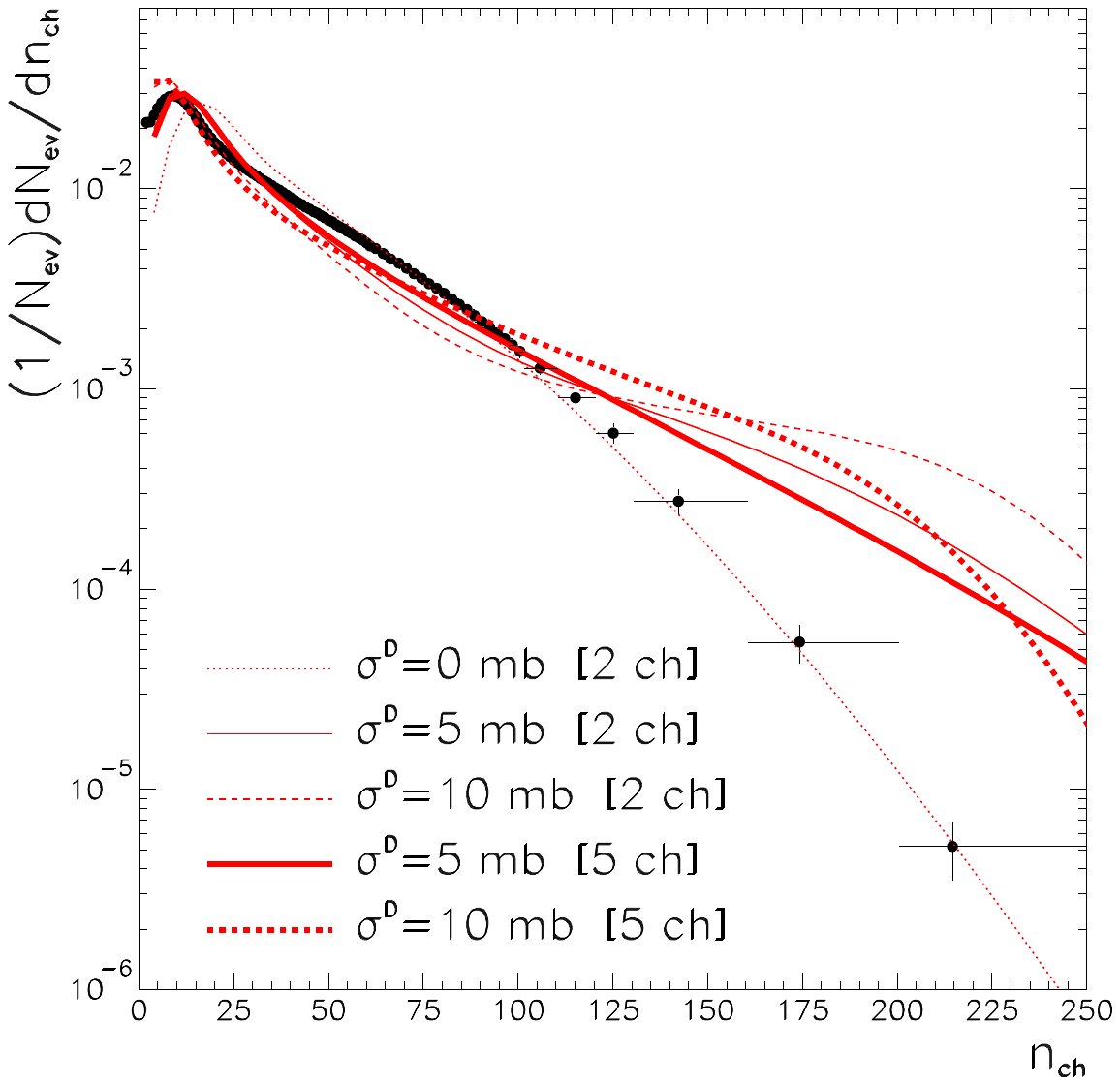}
%\hspace{-0.5cm}
%\includegraphics[scale=0.30]{u-3.pdf}
\vspace{-0.5cm}
\caption{\sf Charged-particle multiplicity distribution in the central region ($-2.5<\eta<2.5$) for the two- and five-channel eikonal models, shown for different values of the low-mass diffractive dissociation cross section. For the two-channel model: no dissociation ($\sigma^D=0$) (thin dotted curve), $\sigma^D=5$ mb (thin solid curve), and $\sigma^D=10$ mb (thin dashed curve). For the five-channel model: $\sigma^D=5$ mb (thick solid curve), and $\sigma^D=10$ mb (thick dashed curve). The data are from~\cite{atl-epg}.
}
\label{f3}
\end{center}
\end{figure*}

In contrast, the distribution is much more sensitive to the dispersion of the couplings $\beta_{i}$, which controls the probability of diffractive dissociation. This effect is illustrated in Fig.~\ref{f3}, where one can see that increasing the value of $\sigma^D$ leads to a more pronounced shoulder in the multiplicity distribution.\footnote{The value $\sigma^D = 5$ mb at 13 TeV is consistent with the 7 TeV TOTEM measurement~\cite{TOT-D} and with the QGSJET-II prediction~\cite{QGSJET}. However, the analysis of Ref.~\cite{GS}, based on a combined study of TOTEM and ATLAS data, suggests that the low-mass diffractive cross section may be as large as $\sim 10$ mb at 13 TeV.}

Finally, we note that, once $\sigma^D$ is fixed, the results depend only weakly on the number of channels $n_{\text{channel}}$, i.e. on the number of Good–Walker eigenstates included in the model. This is demonstrated in Fig.~\ref{f3} by comparing the thin solid ($n_{\text{channel}} = 2$) and thick solid ($n_{\text{channel}} = 5$) curves.\footnote{To obtain $\sigma^D = 5$ mb, we take $\gamma_1 = 1.545$ and $\gamma_2 = 0.455$ in the two-channel case, and $\gamma_1 = 1.73$, $\gamma_2 = 1.295$, $\gamma_3 = 1$, $\gamma_4 = 0.705$, and $\gamma_5 = 0.27$ in the five-channel case.}

\section{Color reconnection and/or the percolation effects}

Recall that when the number of cut Pomerons becomes large (corresponding to a high Pomeron density in $b$-space), their wave functions may overlap. This overlap reduces the effective strength of the sources responsible for the emission of new soft secondary particles. At present, there is no well-established theory that provides a quantitative description of these nonperturbative interactions. Nevertheless, such effects can be modeled qualitatively, for example, by introducing a probability of color reconnection in Monte Carlo generators (see, e.g., \cite{CR1,CR2,CR3}) or by invoking string percolation \cite{BP,BP2,BP3}.

To incorporate these effects in our calculations, we introduce a suppression factor that effectively reduces the number of cut Pomerons $k$ contributing to particle production in the central region: 
\begin{eqnarray}
g(k)=1/(1+\sqrt{k/k_0}) .
\label{fac}
\end{eqnarray}
Thus, the number of produced secondaries is no longer proportional to $k$, but rather to $g(k)k$. For small $k$, one has $g(k)\simeq 1$, and the standard linear scaling is recovered. In contrast, at large $k$, the behavior
\begin{eqnarray}
N_{ch}\propto g(k)k \simeq \sqrt{k}
\end{eqnarray}
emerges, in agreement with expectations based on saturation-like arguments (see, e.g., \cite{BP}).

The parameter $k_0$ controls the onset of the suppression associated with configurations containing a large number of cut Pomerons. In the present
analysis, its value is selected primarily by requiring a reasonable description of the high-multiplicity region of the charged-particle distribution, where the assumption is that all cut Pomerons act as independent sources, leading to a substantial excess. We use $k_0=4.2$ and $k_0=5.2$ for
the two- and five-channel models, respectively. These values should be regarded as limited phenomenological inputs that effectively account for source-overlap effects not explicitly described within the present framework.

As a consistency test of this interpretation, we use the same values of $k_0$ to calculate the production of heavy mesons, in particular $J/\psi$, as a function of the soft charged-particle multiplicity. Since heavy-quark production is expected to be approximately proportional to the number of elementary partonic interactions, whereas the soft multiplicity is reduced by source-overlap effects, the normalized $J/\psi$ yield should increase faster than the normalized charged-particle density. The black curve in Fig.~4 corresponds to the linear-scaling expectation,
\begin{eqnarray}  
\frac{dN^{J/\psi}/dY}{\langle dN^{J/\psi}/dY\rangle}=\frac{dN_{ch}/d\eta}{\langle dN_{ch}/d\eta\rangle} ,
\end{eqnarray}
which underestimates the data. When the suppression factor $g(k)$ is included, the two- and five-channel results give a qualitatively reasonable description of the observed multiplicity dependence \cite{alice}. The $J/\psi$ comparison is therefore not used as an independent determination of $k_0$, but rather as a nontrivial consistency test of the proposed suppression mechanism.

The resulting charged-particle multiplicity distributions at $\sqrt{s}=13$ and $7$ TeV are shown in Figs.~5 and 6, respectively. In both cases, we use $\sigma^{D}=5$ mb and the values of $k_0$ specified above. The suppression substantially reduces the excessively high-multiplicity tail obtained when particle production is assumed to be proportional to the
number of cut Pomerons.

For comparison, the figures also include two complementary descriptions of the same multiplicity data. The first is a phenomenological approach based on the three-component negative-binomial parametrization proposed by Zborovsk\'y. At $7$ TeV, we use the parametrization obtained in Ref. \cite{zboro001}, while the corresponding $13$ TeV result is taken from Ref. \cite{zboro002}. The multiplicity distribution in this approach is represented by a weighted sum of three negative-binomial distributions,
\begin{equation}
P_{\rm 3NBD}(N_{\rm ch})
=
\sum_{i=1}^{3}
\alpha_i
P_{\rm NBD}
\left(
N_{\rm ch};
\left\langle N_{\rm ch}\right\rangle_i,
k_i
\right),
\end{equation}
where the component weights, average multiplicities, and shape parameters are fitted directly to the measured spectra. These curves, therefore, provide flexible phenomenological benchmarks for the detailed shape of the multiplicity distributions.

The second comparison is with QGSJET-II-04. The $7$ TeV result is taken from Ref. \cite{qgsjet001}, while the $13$ TeV curve is the prediction presented by the ATLAS Collaboration in Ref. \cite{qgsjet001}. In contrast to the three-NBD parametrization, QGSJET-II-04 is a complete Monte Carlo event generator that incorporates multiple scattering, multi-Pomeron dynamics, diffraction, energy–momentum conservation, and hadronization. All these comparisons correspond to the event selection correspond to the event selection
\begin{equation}
|\eta|<2.5,
\qquad
p_T>100~{\rm MeV},
\qquad
N_{\rm ch}\geq 2.
\end{equation}

The three descriptions have different objectives and involve different degrees of phenomenological adjustment. The three-NBD result is fitted directly to the full multiplicity spectrum, whereas QGSJET-II-04 contains a detailed event-level model with several phenomenological ingredients. In our calculation, by contrast, the parameters entering the impact-parameter opacity is determined from the forward differential cross section. Once these parameters are fixed, the corresponding total cross sections and the probabilities of configurations containing different numbers of cut Pomerons become predictions of the model.

The dispersion of the Good-Walker couplings is constrained by the assumed value of the low-mass diffractive dissociation cross section, while the average particle yield associated with one cut Pomeron is normalized using the measured inclusive charged-particle density. Thus, apart from the suppression parameter $k_0$, the shape of the multiplicity distribution is largely a prediction of the model and is not obtained through a direct fit to the multiplicity data themselves.

The factor $g(k)$ should not be identified with a specific color-reconnection or string-percolation algorithm. Rather, it provides an inclusive parametrization of the common physical expectation that the number of effectively independent soft-particle-emitting sources grow more slowly than the number of cut Pomerons when their transverse density becomes large. A more microscopic description would require an impact-parameter-dependent treatment of string overlap, color reconnection, or percolation.

Finally, the present model does not include high-mass diffractive dissociation, which is usually described by triple-Pomeron and more complicated enhanced multi-Pomeron diagrams. Secondary particles produced in such processes may fire the edge of central detector and be registered as low-multiplicity events. Their omission is therefore a likely source of the remaining disagreement at small $N_{\rm ch}$. Further improvements would also require a more realistic treatment of the particle production by a single cut Pomeron, including energy-momentum conservation and long-range correlations.

\section{Conclusion}
 
We investigate the multiplicity distribution of charged particles produced at LHC energies in the central rapidity region, $|\eta|<2.5$, within a multi-channel eikonal framework that includes multi-Pomeron exchange. In this approach, particle production is driven by the number of cut Pomerons in a given event, which act as sources of soft secondary particles.

Diffractive effects are implemented through the Good–Walker (GW) formalism, in which the incoming proton is described as a superposition of diffractive eigenstates with different interaction strengths. This allows us to account, in a natural way, for the possibility of low-mass diffractive dissociation. As a result, fluctuations in the interaction strength, encoded in the dispersion of the GW couplings, play an important role in controlling the final-state multiplicity distribution.

We find that the multiplicity distribution is relatively insensitive to the details of short-range rapidity correlations within an individual Pomeron. In particular, modeling particle production as independent emissions, pairs, or clusters yields only minor differences in the final result. Similarly, the dependence on the number of GW eigenstates included in the model is weak, provided that the overall dispersion of the couplings is kept fixed.

In contrast, the distribution shows strong sensitivity to the low-mass diffractive cross section, $\sigma^D$. Physically, this is because $\sigma^D$ controls the degree of fluctuation in the interaction strength from event to event. The presence of GW components with different couplings (i.e., different effective cross sections) leads to a nontrivial structure in the multiplicity distribution. In particular, components with larger cross sections tend to produce more cut Pomerons, thereby increasing the probability of events with high multiplicity. This mechanism naturally generates a characteristic shoulder in the multiplicity distribution at relatively large values of $N_{ch}$.

At very high multiplicities, corresponding to events with a large number of cut Pomerons, the assumption that each Pomeron acts as an independent source of particle production is expected to break down. In such dense environments, the Pomerons overlap in impact-parameter space, and their associated color fields can interact. This leads to collective effects such as color reconnection and/or string percolation, which effectively reduce the number of independent particle-emitting sources and, consequently, suppress the overall particle yield.

To account phenomenologically for the reduction of soft-particle production in configurations containing a large number of cut Pomerons, we introduced the suppression factor $g(k)=1/(1+\sqrt{k/k_0})$. This factor reduces the effective number of independent particle-emitting sources as the Pomeron density increases. The parameter $k_0$ controls when this suppression begins. Its values were chosen to ensure a reasonable fit to the high-multiplicity region of the charged-particle distribution. In the two- and five-channel realizations, we used: 
\begin{equation}
k_0^{(2{\rm ch})}=4.2,
\qquad
k_0^{(5{\rm ch})}=5.2.
\end{equation}
The same values also provide a qualitatively reasonable description of the observed multiplicity dependence of $J/\psi$ production. However, this comparison should not be seen as an independent determination of $k_0$. Instead, it serves as a nontrivial consistency test of interpreting $g(k)$ in terms of source-overlap effects. Examples of such effects include color reconnection, string fusion, or percolation.

To place our results in a broader context, we compared the predicted charged-particle multiplicity distributions at both $\sqrt{s}=7$ and $13$ TeV with two complementary approaches: three-component negative-binomial parametrizations fitted to the measured multiplicity spectra \cite{zboro001,zboro002}, and QGSJET-II-04 \cite{qgsjet001,qgsjet002}, an event-level generator incorporating multiple scattering, multi-Pomeron dynamics, diffraction, energy-momentum conservation, and hadronization.

These comparisons highlight the different purposes of our calculations. The three-NBD parametrization is meant to reproduce the detailed shape of the measured distribution. In contrast, QGSJET-II-04 gives a full Monte Carlo description of the final state. Our approach aims to preserve the analytic relation between the opacity, the Good-Walker fluctuations, probabilities for different numbers of cut Pomerons, and the resulting multiplicity distribution. In this way, the formation of the shoulder and the suppression of the high-multiplicity tail can be linked directly to the underlying physical mechanisms.

We emphasize the limited tuning required for this calculation. The parameters of the impact-parameter opacity are determined only from fits to the forward differential elastic cross-section data. Once these parameters are fixed, both the total cross sections and the probabilities of configurations with different numbers of cut Pomerons become predictions. The dispersion of the Good-Walker couplings is constrained by the assumed low-mass diffractive dissociation cross section. The average particle yield associated with one cut Pomeron is normalized using the measured inclusive charged-particle density. Apart from the suppression parameter $k_0$, most features of the multiplicity distribution are thus controlled by information not taken from a direct fit to its detailed shape.

The remaining discrepancies suggest concrete avenues for improvement. The low-multiplicity region likely requires high-mass diffractive dissociation, which is absent here and would need triple-Pomeron and enhanced multi-Pomeron diagrams. Achieving a more quantitative description would also call for a richer Good-Walker configuration representation and a realistic treatment of particle production by individual cut Pomerons, incorporating energy-momentum conservation, resonance decays, fluctuations, and long-range correlations.

Finally, the global factor $g(k)$ should eventually be replaced by a more microscopic and impact-parameter-dependent description of string overlap, color reconnection, or percolation. Such developments would preserve the analytic transparency of the present approach while incorporating some of the dynamical ingredients currently treated only in complete Monte Carlo event generators.

%To model this effect, we introduce a phenomenological suppression factor that reduces the effective number of particle-producing sources at large Pomeron multiplicities. The form of this factor is constrained using experimental data on $J/\psi$ production as a function of the charged-particle multiplicity, which provides a sensitive probe of the underlying dynamics.

Our final results, including these suppression effects, are compared with ATLAS data at $\sqrt{s}=7$ \cite{atlas-epg7} and 13 \cite{atl-epg} TeV, showing a good overall agreement in the high multiplicity region.

\section*{Acknowledgments}

This research was partially supported by the Conselho Nacional de Desenvolvimento Cient\'{\i}fico e Tecnol\'ogico (CNPq).
 
%The authors thank V.A. Khoze for reading the manuscript and for valuable discussions.

\begin{figure*}[t]
\begin{center}
\vspace{-3.5cm}
\includegraphics[scale=0.6]{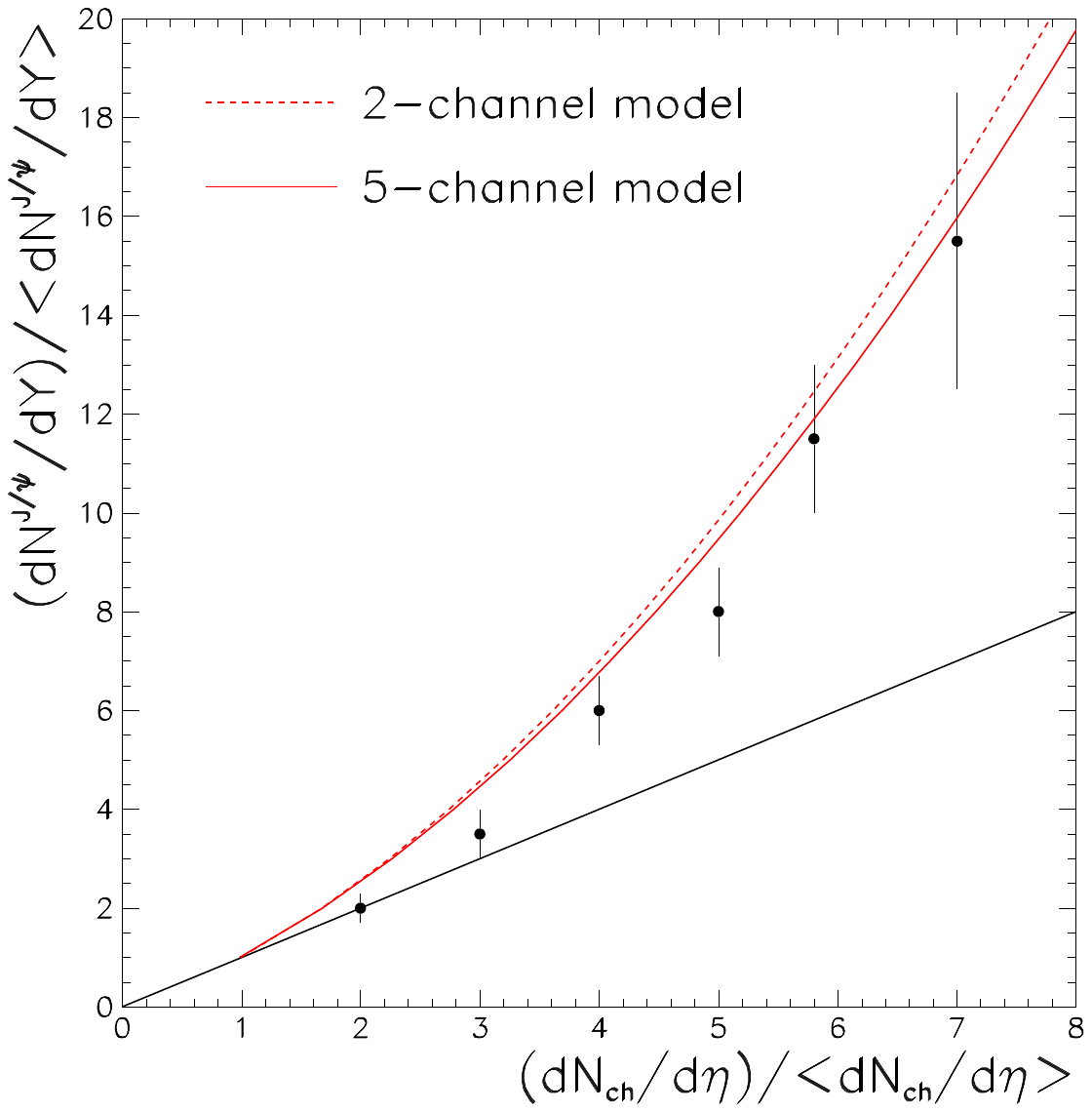}
%\hspace{-0.5cm}
%\includegraphics[scale=0.30]{u-3.pdf}
\vspace{-0.5cm}
\caption{\sf Yield of $J/\psi$ mesons in the central region ($|Y|<0.9$) depending on the central ($|\eta|<1$) multiplicity in the event. The curves include the suppression factor (\ref{fac}) (with $k_0=4.2$ for the two-channel model and $k_0=5.2$ for the five-channel model) of soft secondaries emission from the configurations with a large number of Pomerons (i.e., high Pomeron density). The data are from~\cite{alice}.
}
\label{f4}
\end{center}
\end{figure*}

\begin{figure*}[t]
\begin{center}
\vspace{-3.5cm}
\includegraphics[scale=0.6]{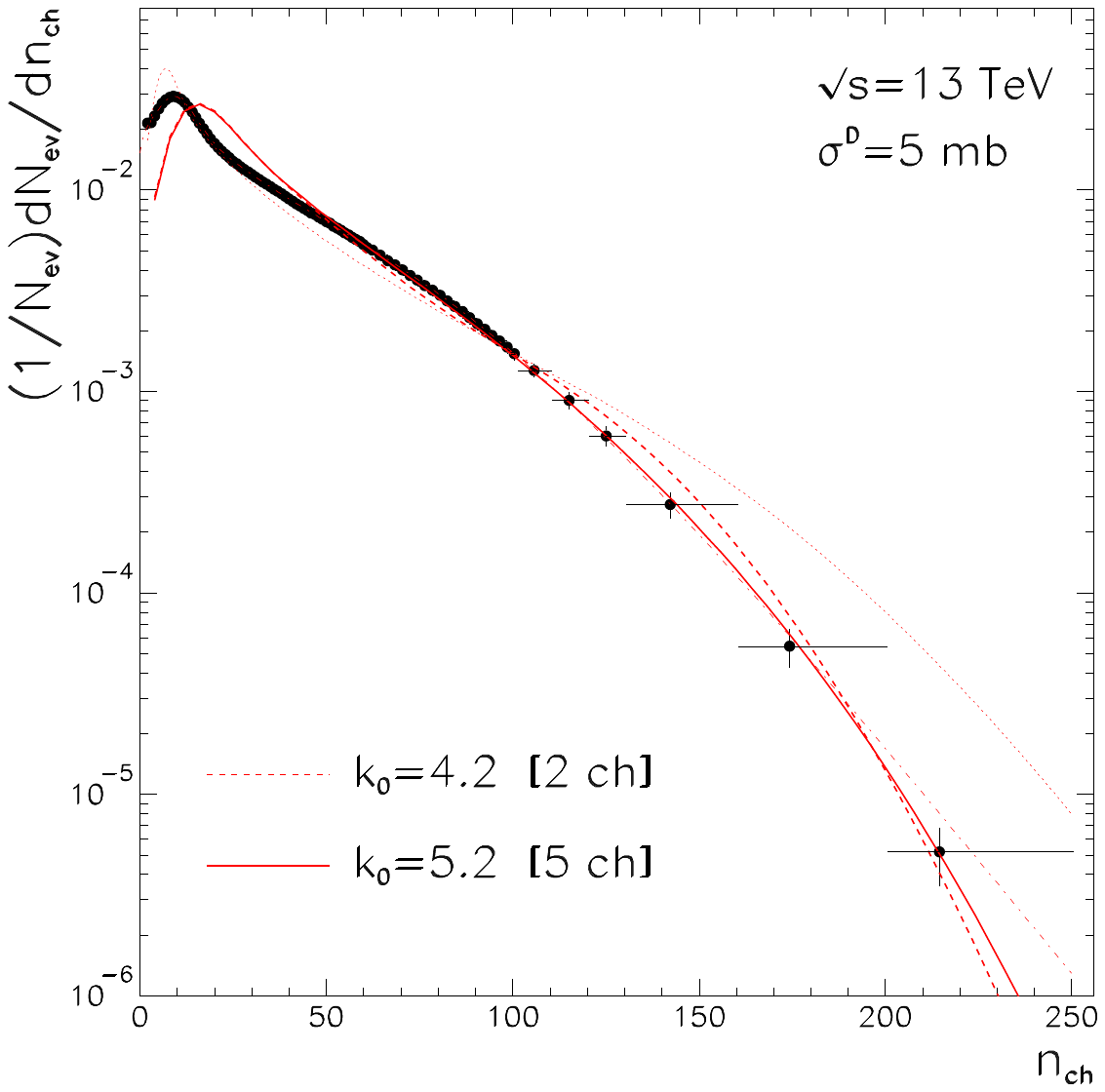}
%\hspace{-0.5cm}
%\includegraphics[scale=0.30]{u-3.pdf}
\vspace{-0.5cm}
\caption{\sf Charged-particle multiplicity distribution in the central region ($-2.5<\eta<2.5$) at $\sqrt s=13$ TeV, for the two-channel (dashed curve) and the five-channel (solid curve) eikonal models. The results include the suppression factor (\ref{fac}), accounting for the reduced emission of soft secondaries in configurations with a large number of Pomerons. The dotted curve represents the QGSJET-II-04 result, while the dash-dotted curve corresponds to the three-NBD parametrization. The data are from~\cite{atl-epg}.
}
\label{f5}
\end{center}
\end{figure*}

\begin{figure*}[t]
\begin{center}
\vspace{-3.5cm}
\includegraphics[scale=0.6]{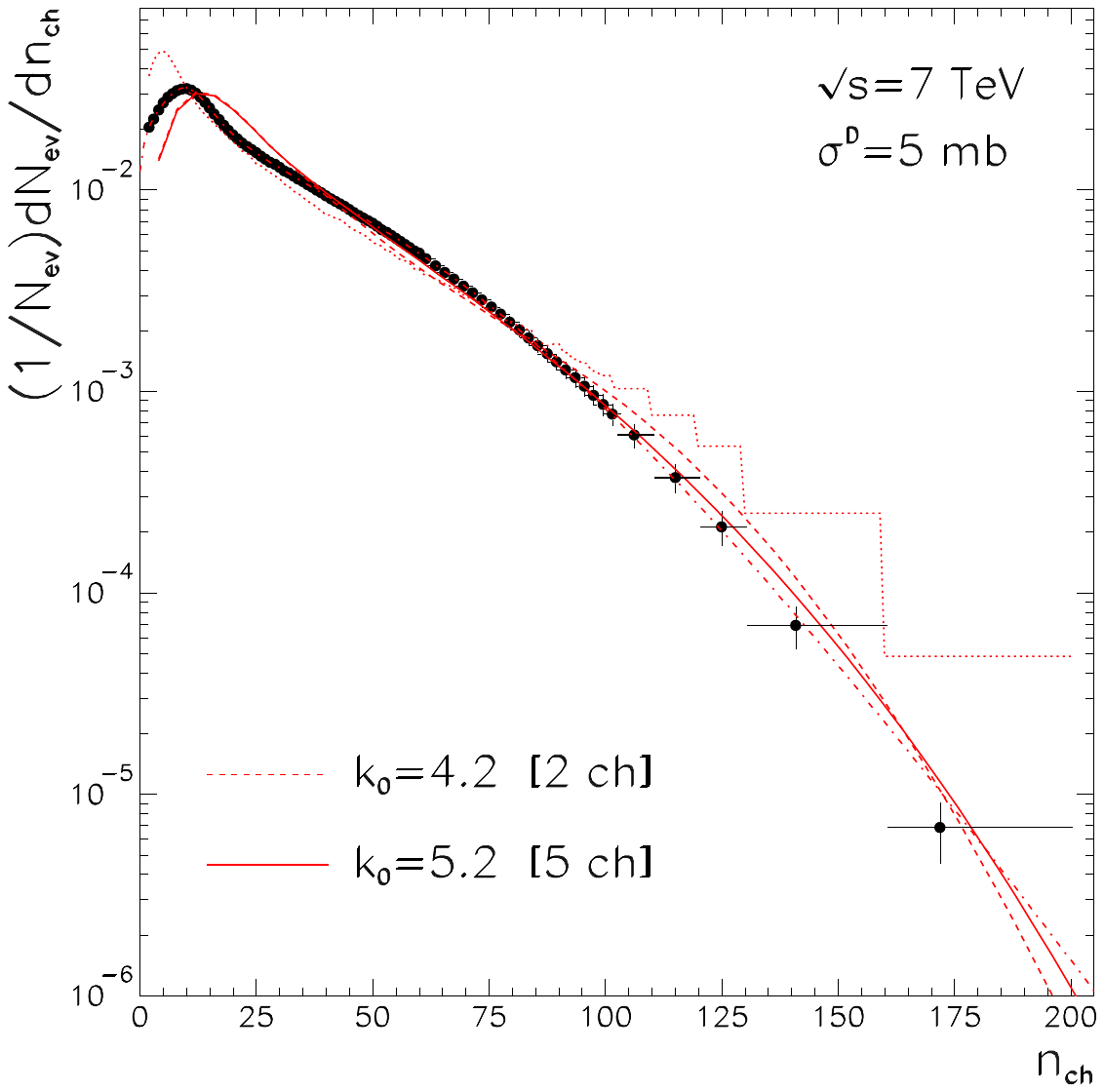}
%\hspace{-0.5cm}
%\includegraphics[scale=0.30]{u-3.pdf}
\vspace{-0.5cm}
\caption{\sf Charged-particle multiplicity distribution in the central region ($-2.5<\eta<2.5$) at $\sqrt s=7$ TeV, for the two-channel (dashed curve) and the five-channel (solid curve) eikonal models. The results include the suppression factor (\ref{fac}), accounting for the reduced emission of soft secondaries in configurations with a large number of Pomerons. The dotted curve represents the QGSJET-II-04 result, while the dash-dotted curve corresponds to the three-NBD parametrization. The data are from~\cite{atlas-epg7}.
}
\label{f6}
\end{center}
\end{figure*}

\end{document}